\documentclass[journal]{IEEEtran}

\usepackage{ifthen}
\usepackage[cmex10]{amsmath}
\usepackage{color}
\usepackage{subfigure}
\usepackage{amsmath}
\usepackage{mathtools}
\usepackage{amssymb}
\usepackage{marvosym}
\usepackage{type1cm}
\usepackage{epic}
\usepackage{wasysym}
\usepackage{latexsym}
\usepackage{graphicx}
\usepackage{placeins}
\usepackage{subfigure}
\usepackage{url}
\usepackage{fancyhdr}
\usepackage{wrapfig}
\usepackage{cite}
\usepackage{url}
\usepackage{psfrag}
\usepackage{enumerate}
\newcommand{\set}[1]{\mathcal{#1}}   
\newcommand{\cset}[1]{\mathcal{#1}^{\textnormal{c}}} 
\newcommand{\wt}[1]{\widetilde{#1}}   

\newtheorem{theorem}{Theorem}
\newtheorem{example}{Example}
\newtheorem{remark}{Remark}
\newtheorem{lemma}{Lemma}

\begin{document}

\title{Effective Secrecy: Reliability, Confusion and Stealth}

 \IEEEoverridecommandlockouts

\author{
\IEEEauthorblockN{Jie Hou and Gerhard Kramer}\\
\IEEEauthorblockA{Institute for Communications Engineering\\
   Technische Universit\"at M\"unchen, 80290 Munich, Germany\\
       Email: \{jie.hou, gerhard.kramer\}@tum.de}
}

\maketitle

\begin{abstract}
A security measure called effective security is defined that includes strong secrecy and stealth communication. Effective secrecy 
ensures that a message cannot be deciphered and that the presence of meaningful communication is hidden. To measure stealth we 
use resolvability and relate this to binary hypothesis testing. Results are developed for wire-tap channels and broadcast 
channels with confidential messages.
\end{abstract}

\section{Introduction}

Wyner \cite{Wyner03} derived the {\em secrecy capacity} for {\em degraded} {wire-tap channels} (see Fig.~\ref{wiretap}).
Csisz{\'a}r and K{\"o}rner \cite{Csiszar02} extended the results to broadcast channels with confidential messages. In both
\cite{Wyner03} and \cite{Csiszar02}, secrecy was measured by a {\em normalized} mutual information between the message $M$ and the
eavesdropper's output $Z^n$ under a secrecy constraint
\begin{align}
\frac{1}{n}I(M;Z^n)\le S
\end{align}
which is referred to as {\em weak secrecy}. Weak secrecy has the advantage that one can trade off $S$ for rate. The drawback is 
that even $S\approx 0$ is usually considered too weak because the eavesdropper can decipher $nS$ bits of $M$, which grows 
with $n$. Therefore, \cite{Maurer01} (see also \cite{Csiszar03}) advocated using {\em strong secrecy} where secrecy
is measured by the {\em unnormalized} mutual information $I(M;Z^n)$ and requires
\begin{align}
I(M;Z^n) \le \xi
\end{align}
for any $\xi >0$ and sufficiently large $n$.

In related work, Han and Verd\'{u} \cite{Verdu01} studied {\em resolvability} based on {\em variational distance} that addresses the
number of bits needed to mimic a marginal distribution of a prescribed joint distribution. Bloch and Laneman \cite{Bloch02} used
the resolvability approach of \cite{Verdu01} and extended the results in \cite{Csiszar02} to continuous random variables and
channels with memory.

The main contribution of this work is to define a new and stronger security measure for wire-tap channels that includes not only 
reliability and (wiretapper) confusion but also {\em stealth}. The measure is satisfied by random codes and by using a recently 
developed simplified proof \cite{Hou01} of resolvability based on {\em unnormalized} informational divergence (see 
also \cite[Lemma 11]{Winter02}). In particular, we measure secrecy by the informational divergence
\begin{align}
D(P_{MZ^n}||P_MQ^n_Z)
\end{align}
where $P_{MZ^n}$ is the joint distribution of $MZ^n$, $P_M$ is the
distribution of $M$, $P_{Z^n}$ is the distribution of $Z^n$, and $Q^n_Z$ is the distribution
that the eavesdropper expects to observe when the source is {\em not} communicating useful messages. We call this security measure 
{\em effective secrecy}. One can easily check that (see (\ref{twoTerm}) below)
\begin{align}
&D(P_{MZ^n}||P_MQ^n_Z)=\underset{\text{Non-Confusion}}{\underbrace{I(M;Z^n)}}+\underset{\text{Non-Stealth}}{\underbrace{D(P_{Z^n}
||Q^n_Z)}}\label{twoguys}
\end{align}
where we interpret $I(M;Z^n)$ as a measure of \lq\lq non-confusion\rq\rq\;and $D(P_{Z^n}||Q^n_Z)$ as a measure of \lq\lq 
non-stealth\rq\rq. We justify the former interpretation by using error probability in Sec. \ref{mainResult} and the latter by 
using binary hypothesis testing in Sec. \ref{Hypo}. Thus, by making $D(P_{MZ^n}||P_MQ^n_Z)\rightarrow 0$ we not only keep the 
message secret from the eavesdropper but also hide the presence of meaningful communication. 


The paper is organized as follows. In Section \ref{notation}, we state the problem. In Section \ref{mainResult} we state and
prove the main result. Section \ref{Hypo} relates the result to hypothesis testing. Section \ref{Diss} discusses related works.

\section{Preliminaries}
\label{notation}

\subsection{Notation}
Random variables are written with upper case letters and their realizations with the corresponding lower
case letters. Superscripts denote finite-length sequences of variables/symbols, e.g., $X^n=X_{1},\dots, X_{n}$. Subscripts
denote the position of a variable/symbol in a sequence. For instance, $X_{i}$ denotes the $i$-th variable in $X^n$. We use
$X^n_i$ to denote the sequence $X_i,\dots,X_n$, $1\le i \le n$. A random variable $X$ has probability distribution $P_X$ and the
support of $P_X$ is denoted as $\text{supp}(P_X)$. We write probabilities with subscripts $P_{X}(x)$ but we drop the subscripts 
if the arguments of the distribution are lower case versions of the random variables. For example, we write $P(x)=P_X(x)$. If the 
$X_i$, $i=1,\dots,n$, are independent and identically distributed (i.i.d.) according to $P_X$, then we have
$P(x^n)=\prod^n_{i=1}P_X(x_i)$ and we write $P_{X^n}=P^n_X$. We often also use $Q^n_X$ to refer
to sequences of i.i.d. random variables. Calligraphic letters denote sets. The size of a set $\set S$ is denoted as $|\set S|$
and the complement is denoted as $\cset S$. For $X$ with alphabet $\set X$, we denote $P_X(\set S)=\sum_{x\in \set S} P_X(x)$ 
for any $\set S\subseteq \set X$. We use $\set T^n_\epsilon(P_X)$ to denote the set of letter-typical sequences of length $n$ with
respect to the probability distribution $P_X$ and the non-negative number $\epsilon$ \cite[Ch.
3]{ Massey01},
\cite{Roche01 }, i.e., we have
\begin{align*}
\set T^n_\epsilon(P_X)=\left\{x^n:\Big| \frac{N(a|x^n)}{n} -P_X(a) \Big| \le \epsilon P_X(a),\; \forall a\in \set X  \right\}
\end{align*}
where $N(a|x^n)$ is the number of occurrences of $a$ in $x^n$.

\subsection{Wire-Tap Channel}

\begin{figure}[t!]
\centering
\psfrag{alice}[][][1]{\text{Joey}}
\psfrag{bob}[][][1]{\text{Chandler}}
\psfrag{eve}[][][1]{\text{Ross}}
\psfrag{M0}[][][1]{$M$}
\psfrag{W}[][][1]{$W$}
\psfrag{unW}[][][1]{$U^n$}
\psfrag{Xn}[][][1]{$X^n$}
\psfrag{Yn}[][][1]{$Y^n$}
\psfrag{Zn}[][][1]{$Z^n$}
\psfrag{hatM0}[][][1]{$\hat M$}
\psfrag{hatM2}[][][1]{$\hat W$}
\psfrag{Pxyz}[][][1]{$Q^n_{YZ|X}$}
\psfrag{f}[][][1]{Encoder}
\psfrag{d1}[][][1]{Decoder}
\psfrag{info}[][][1]{$I(M; Z^n)$}
\psfrag{Dive}[][][1]{$D(P_{Z^n}||Q^n_Z)$}
\includegraphics[width=8.87cm]{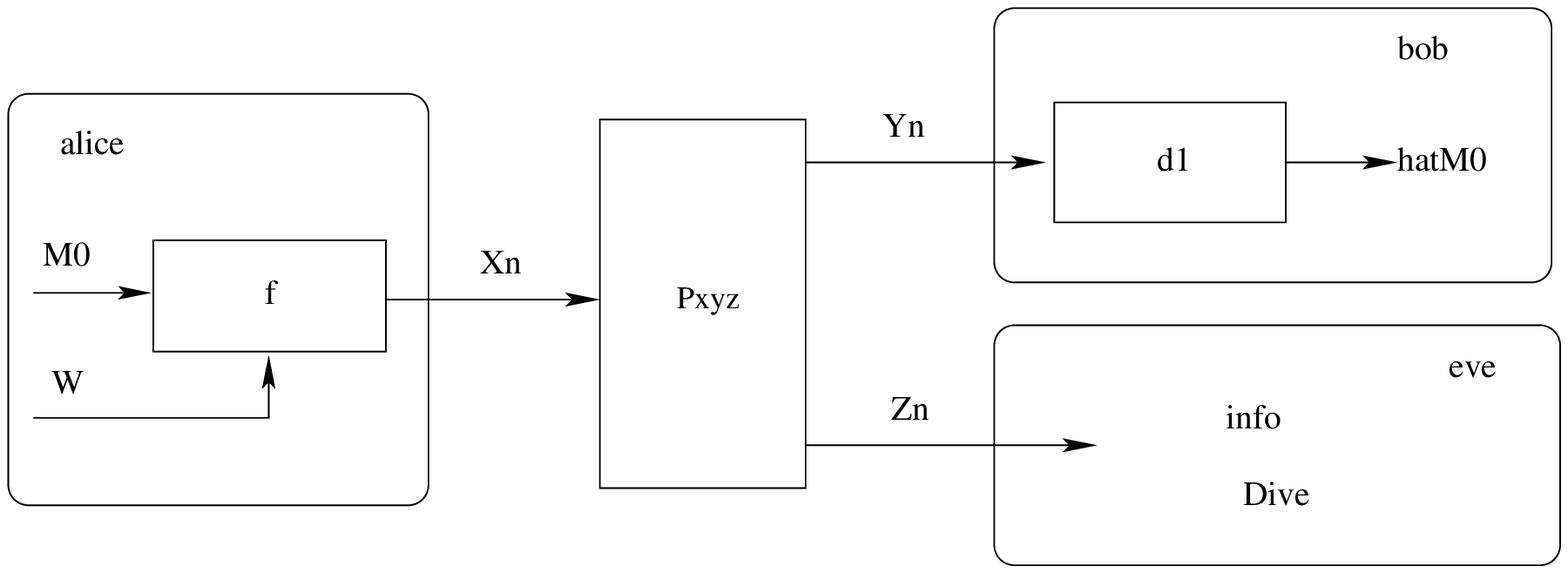}
\caption{A wire-tap channel.}
\label{wiretap}
\end{figure}
Consider the wire-tap channel depicted in Fig. \ref{wiretap}. Joey has a message $M$ which is destined for Chandler but
should be kept secret from Ross. The message $M$ is uniformly distributed over $\{1,\dots, L\}$, $L=2^{nR}$, and an encoder
$f(\cdot)$ maps $M$ to the sequence
\begin{align}
X^n=f(M,W)
\end{align}
with help of a randomizer variable $W$ that is independent of $M$ and uniformly distributed over $\{1,\dots,
L_1\}$, $L_1=2^{nR_1}$. The purpose of $W$ is to confuse Ross so that he learns little about $M$. $X^n$ is transmitted through a
memoryless channel $Q^n_{YZ|X}$. Chandler observes the channel output $Y^n$ while Ross observes $Z^n$. The pair $MZ^n$ has the
joint distribution $P_{MZ^n}$. Chandler estimates $\hat M$ from $Y^n$ and the average error probability is
\begin{align}
P^{(n)}_e&=\Pr\left[\hat M \ne M\right]\label{errorM}.
\end{align}
Ross tries to learn $M$ from $Z^n$ and secrecy is measured by
\begin{align}
&D(P_{MZ^n}||P_M Q^n_Z)\notag \\
&=\sum_{ \underset{\in \text{supp}(P_{MZ^n})}{(m, z^n)}} P(m,z^n) \log\left(\frac{P(m, z^n)}{P(m)\cdot Q^n_Z(z^n)} \cdot
\frac{P(z^n)}{P(z^n) } \right)\notag\\
&=\sum_{ \underset{\in \text{supp}(P_{MZ^n})}{(m, z^n)}} P(m,z^n)
\left(\log\frac{P(z^n|m)}{P(z^n)}+\log\frac{P(z^n)}{Q^n_Z(z^n)}\right)\notag\\
&=\underset{\text{Non-Confusion}}{\underbrace{I(M;Z^n)}}+\underset{\text{Non-Stealth}}{\underbrace{D(P_{Z^n}||Q^n_Z)}}
\label{twoTerm}
\end{align}
where $P_{Z^n}$ is the distribution Ross observes at his channel output and $Q^n_Z$ is the distribution Ross expects
to observe if Joey is {\em not} sending useful information. For example, if Joey transmits $X^n$ with probability
$Q^n_X(X^n)$ through the channel, then we have
\begin{align}
Q^n_Z(z^n)=\sum_{x^n\in \text{supp}(Q^n_X)}Q^n_X(x^n)Q^n_{Z|X}(z^n|x^n)\label{Qzn}.
\end{align}
When Joey sends useful messages, then $P_{Z^n}$ and $Q^n_Z$ are different. But a small
$D(P_{MZ^n}||P_MQ^n_Z)$ implies that both $I(M;Z^n)$ and $D(P_{Z^n}||Q^n_Z)$ are small which in turn
implies that Ross learns little about $M$ and cannot recognize whether Joey is communicating anything meaningful.
A rate $R$ is {\em achievable} if for any $\xi_1, \xi_2 >0$ there is a sufficiently large $n$ and an encoder and a decoder such 
that
\begin{align}
 P^{(n)}_e& \le \xi_1 \label{reliability}\\
D(P_{MZ^n}||P_M Q^n_Z)&\le \xi_2 \label{stealth}.
\end{align}
The {\em effective secrecy capacity} $C_S$ is the supremum of the set of achievable $R$. We wish to determine $C_S$.

\section{Main result and Proof}
\label{mainResult}
We prove the following result.
\begin{theorem}
\label{thm1}
The effective secrecy capacity of the wire-tap channel is the same as the weak and strong secrecy capacity, namely
\begin{align}
C_S=\underset{Q_{VX}}{\text{max}}\; \left[ I(V;Y)-I(V;Z) \right]
\end{align}
where the maximization is over all joint distributions $Q_{VX}$ satisfying the Markov chain
\begin{align}
&V-X-YZ \label{markov}.
\end{align}
One may restrict the cardinality of $V$ to $|\set V| \le |\set X|$ \label{cardiBound}.

\end{theorem}

\subsection{Achievability}
\label{Achievability}
We use random coding and the proof technique of \cite{Hou01}.

{\em Random Code:} Fix a distribution $Q_X$ and generate $L\cdot L_1$ codewords $x^n(m, w)$, $m=1,\dots, L$,
$w=1,\dots, L_1$ using $\prod^n_{i=1}Q_X(x_i(m,w))$. This defines the codebook
\begin{align}
\set C=\{ x^n(m, w), m=1,\dots, L,\; w=1,\dots, L_1\}
\end{align}
and we denote the random codebook by
\begin{align}
\set {\wt C}=\{X^n(m, w)\}^{(L,L_1)}_{(m,w)=(1,1)}.
\end{align}

{\em Encoding:} To send a message $m$, Joey chooses $w$ uniformly from $\{1,\dots, L_1\}$ and transmits $x^n(m,w)$.
Hence, for a fixed codebook $\set C$ every $x^n(m,w)$ occurs with probability
\begin{align}
P_{X^n}(x^n(m,w))=\frac{1}{L\cdot L_1}
\end{align}
rather than $Q^n_X(x^n(m,w))$. Further, for every pair $(m, z^n)$ we have (see (\ref{Qzn}))
\begin{align}
P(z^n|m)&=\sum^{L_1}_{w=1} \frac{1}{L_1}\cdot Q^n_{Z|X}(z^n|x^n(m,w))\label{condiProb}\\
P(z^n)&=\sum^{L}_{m=1}\sum^{L_1}_{w=1} \frac{1}{L\cdot L_1}\cdot Q^n_{Z|X}(z^n|x^n(m,w))\label{margiProb}.
\end{align}

{\em Chandler:} Chandler puts out $(\hat m, \hat w)$ if there is a unique pair $(\hat m,\hat w)$ satisfying the
typicality check
\begin{align}
(x^n(\hat m, \hat w), y^n)\in \set T^n_\epsilon(Q_{XY}).
\end{align}
Otherwise he puts out $(\hat m, \hat w)=(1,1)$.

{\em Analysis:} Define the events
\begin{align}
E_1&: \{(\hat M,\hat W)\ne (M, W )\}\notag\\
E_2 &: D(P_{MZ^n}||P_M Q^n_Z)> \xi_2.
\end{align}
Let $E= E_1\cup E_2$ so that we have
\begin{align}
\Pr[E]\le \Pr[E_1] + \Pr[E_2]\label{union}
\end{align}
where we have used the union bound. $\Pr[E_1]$ can be made small with large $n$ as long as
\begin{align}
R+R_1<I(X;Y)-\delta_\epsilon(n) \label{goodcode1}
\end{align}
where $\delta_\epsilon(n)\rightarrow 0$ as $n\rightarrow \infty$ (see \cite{Roche01}) which implies that $P^{(n)}_e$ is
small.

$\Pr[E_2]$ can be made small with large $n$ as long as \cite[Theorem 1]{Hou01}
\begin{align}
R_1>I(X;Z)+\delta^\prime_\epsilon(n) \label{goodcode2}
\end{align}
where $\delta^\prime_\epsilon(n)\rightarrow 0$ as $n\rightarrow \infty$. This is because the average divergence over $M$, $W$,
$\set {\wt C}$ and $Z^n$ satisfies (see \cite[Equ. (9)]{Hou01})
\begin{align}
&\text{E}[D(P_{MZ^n}||P_MQ^n_Z)]\notag\\
 &\overset{(a)}{=}\text{E}[ D(P_{M}||P_{M})+ D(P_{Z^n|M}||Q^n_Z|P_M)]\notag\\
&\overset{(b)}{=} \text{E}\left[\log\frac{\sum^{L_1}_{j=1}Q^n_{Z|X}(Z^n|X^n(M,j))}{L_1\cdot Q^n_Z(Z^n)}\right]\notag\\
&=\sum^L_{m=1}\sum^{L_1}_{w=1}\frac{1}{L\cdot L_1}\notag\\
&\quad \text{E}\left[\log\frac{\sum^{L_1}_{j=1}Q^n_{Z|X}(Z^n|X^n(m,j))}{L_1\cdot Q^n_Z(Z^n)}\Bigg|M=m,W=w\right]\notag\\
&\overset{(c)}{\le}\sum^L_{m=1}\sum^{L_1}_{w=1} \frac{1}{L\cdot L_1} \notag\\
&\quad\text{E}\left[\log\left(\frac{Q^n_{Z|X}(Z^n|X^n(m,w)) }{L_1\cdot Q^n_Z(Z^n)} +1\right)\Bigg|M=m,W=w\right]\notag\\
&\overset{(d)}{=}\text{E} \left[\log\left(\frac{Q^n_{Z|X}(Z^n|X^n) }{L_1\cdot Q^n_Z(Z^n)} +1\right)\right]\label{avediv}
\end{align}
where
\begin{enumerate}[(a)]
\item follows from the chain rule for informational divergence;
 \item follows from (\ref{condiProb}) and by taking the expectation over $M, W, X^n(1,1),\dots, X^n(L,L_1), Z^n$;
\item follows by the concavity of the logarithm and Jensen's inequality applied to the expectation over the $X^n(m, j), j\ne w$
for a fixed $m$;
\item follows by choosing $X^nZ^n\sim Q^n_{XZ}$.
\end{enumerate}
Next we can show that the right hand side (RHS) of (\ref{avediv}) is small if (\ref{goodcode2}) is valid by splitting the 
expectation in (\ref{avediv}) into sums of typical and atypical pairs (see \cite[Equ. (11)-(16)]{Hou01}). But if the RHS of 
(\ref{avediv}) approaches $0$, then using (\ref{twoTerm}) we have
\begin{align}
\text{E}\left[ I(M;Z^n) + D(P_{Z^n}||Q^n_Z) \right]\rightarrow 0.
\end{align}
Combining (\ref{union}), (\ref{goodcode1}) and (\ref{goodcode2}) we can make $\Pr[E]\rightarrow 0$ as $n\rightarrow \infty$ as 
long as
\begin{align}
R+R_1&<I(X;Y)\\
R_1&>I(X;Z)\label{goodcode22}.
\end{align}
We hence have the achievability of any $R$ satisfying
\begin{align}
0\le R&<\underset{Q_{X}}{\text{max}}\;[I(X;Y)-I(X;Z)]. \label{maxX}
\end{align}
Of course, if the RHS of (\ref{maxX}) is non-positive,
then we require $R=0$. Now we prefix a channel $Q^n_{X|V}$ to the original channel $Q^n_{YZ|X}$ and obtain a new channel
$Q^n_{YZ|V}$ where
\begin{align}
&Q^n_{YZ|V}(y^n,z^n|v^n)\notag\\
&=\sum_{x^n\in \text{supp}(Q^n_{X|V}(\cdot|v^n))}Q^n_{X|V}(x^n|v^n)Q^n_{YZ|X}(y^n,z^n|x^n).
\end{align}
Using a similar analysis as above, we have the achievability of any $R$ satisfying
\begin{align}
0\le R&<\underset{Q_{VX}}{\text{max}}\;[I(V;Y)-I(V;Z)]\label{maxV}
\end{align}
where the maximization is over all $Q_{VX}$ satisfying (\ref{markov}). Again, if the RHS of
(\ref{maxV}) is non-positive, then we require $R=0$. As usual, the purpose of adding the auxiliary variable $V$ is to potentially 
increase $R$. Note that $V=X$ recovers (\ref{maxX}). Hence, the RHS of (\ref{maxX}) is always smaller than or equal to the RHS of 
(\ref{maxV}).
\begin{remark}
The average divergence $\text{E}[D(P_{MZ^n}||P_MQ^n_Z)]$ can be viewed as the sum of $I(M\set {\wt C}; Z^n)$ and
$D(P_{Z^n}||Q^n_Z)$ \cite[Sec. III]{Ha01} (see also
\cite[Sec. III-B]{Hou01}). To see this, consider
\begin{align}
&\text{E}[D(P_{MZ^n}||P_MQ^n_Z)]\notag\\
&\overset{(a)}{=}\text{E}\left[ \log\frac{\sum^{L_1}_{j=1}Q^n_{Z|X}(Z^n|X^n(M,j)) }{L_1\cdot Q^n_Z(Z^n)}\right]\notag\\
&=\sum^L_{m=1}\frac{1}{L}\sum_{x^n(1,1)}\cdots \sum_{x^n(L, L_1)}\prod^{(L, L_1)}_{k=(1,1)}Q^n_X(x^n(k))\notag\\
&\quad\quad \sum_{z^n}\sum^{L_1}_{w=1}\frac{1}{
L_1}Q^n_{Z|X}(z^n|x^n(m,w))\notag\\
&\quad\quad \log\left[\frac{\sum^{L_1}_{j=1}\frac{1}{L_1}Q^n_{Z|X}(z^n|x^n(m,j))
}{Q^n_Z(z^n)}\right]\notag\\
&=\sum^L_{m=1}P(m)\sum_{\set C}P(\set C|m)\sum_{z^n}P(z^n|m,\set C) \log\frac{P(z^n|m, \set C)}{Q^n_Z( z^n)}\notag\\
&=\sum_{(m, \set C, z^n)} P(m, \set C, z^n) \left(\log\frac{P(z^n|m, \set C)}{P(z^n)} + \log\frac{P(z^n)}{Q^n_{Z}(z^n)}
\right)\notag\\
&=I(M\set{\wt C};Z^n) +D(P_{Z^n}||Q^n_Z)
\end{align}
where (a) follows by~(\ref{avediv})(b). Therefore, as $\text{E}[D(P_{MZ^n}||P_MQ^n_Z)]\rightarrow 0$ we have $I(M\set{\wt C}; 
Z^n) \rightarrow 0$ which means that $M\set {\wt C}$ and $Z^n$ are (almost) independent. This makes sense, since for effective 
secrecy the adversary learns little about $M$ nor about the presence of meaningful transmission.
\end{remark}

\subsection{Converse}
The converse follows as in \cite[Theorem 1]{Csiszar02}. We provide an alternative proof using the {\em telescoping 
identity} \cite[Sec. G]{Kramer01}. Suppose that for some $\xi_1$, $\xi_2>0$ there exists a sufficiently large $n$, an encoder and 
a decoder such that (\ref{reliability}) and (\ref{stealth}) are satisfied.
We have
\begin{align}
&\log_2 L=nR\notag\\
&=H(M)\notag\\
&=I(M;Y^n)+H(M|Y^n)\notag\\
&\overset{(a)}{\le} I(M;Y^n)+\left( 1+\xi_1\cdot nR \right)\notag\\
&\overset{(b)}{\le} I(M;Y^n)-I(M;Z^n)+\xi_2+\left(1+\xi_1\cdot nR\right)\label{equ1}
\end{align}
where $(a)$ follows from Fano's inequality and $(b)$ follows from (\ref{twoTerm}) and (\ref{stealth}). Using the telescoping identity \cite[Equ. (9) and (11)]{Kramer01} we have
\begin{align}
&\frac{1}{n}\left [I(M;Y^n)-I(M;Z^n) \right]\notag\\
&=\sum^n_{i=1}[I(M;Z^n_{i+1}Y^i)-I(M;Z^n_iY^{i-1}) ]\notag\\
&=\frac{1}{n}\sum^n_{i=1}[I(M;Y_i|Y^{i-1}Z^{n}_{i+1} ) -I(M; Z_{i};| Y^{i-1}Z^n_{i+1})]\notag\\
&\overset{(a)}{=} I(M;Y_T|Y^{T-1}Z^n_{T+1}T )-I(M;Z_T|Y^{T-1}Z^n_{T+1}T)\notag\\
&\overset{(b)}{=} I(V;Y|U)-I(V;Z|U)\notag\\
&\le \underset{Q_{UVX}}{\text{max}}\;[I(V;Y|U)-I(V;Z|U)]\notag\\
&\le \underset{u}{\text{max}}\; \underset{Q_{VX|U=u}}{\text{max}}\;[I(V;Y|U=u)-I(V;Z|U=u)]\label{equ3}\\
&\overset{(c)}{=}\underset{Q_{VX}}{\text{max}}\;[I(V;Y)-I(V;Z)] \label{equ2}
\end{align}
where
\begin{enumerate}[(a)]
\item follows by letting $T$ be independent of all other random variables and uniformly distributed over
$\{1,\dots,n\}$;
\item follows by defining
\begin{align}
U&=Y^{T-1}Z^n_{T+1}T,\;V=MU,\notag\\
X&=X_T, \; Y=Y_T, \;Z=Z_T;
\end{align}
\item follows because if the maximum in (\ref{equ3}) is achieved for $U=u^\ast$ and $Q_{VX|U=u^\ast}$, then the same can be
achieved in (\ref{equ2}) by choosing a $Q_{VX}=Q_{VX|U=u^\ast}$.
\end{enumerate}
Combining (\ref{equ1}) and (\ref{equ2}) we have
\begin{align}
R&\le \frac{\underset{Q_{VX}}{\text{max}}\;[I(V;Y)-I(V;Z)]}{1-\xi_1}+\frac{\xi_2+1}{(1-\xi_1)n}.
\end{align}
Letting $n\rightarrow \infty$, $\xi_1\rightarrow 0$, and $\xi_2 \rightarrow 0$, we have
\begin{align}
R\le \underset{Q_{VX}}{\text{max}}\;[I(V;Y)-I(V;Z)]
\end{align}
where the maximization is over all $Q_{VX}$ satisfying the Markov chain (\ref{markov}). The
cardinality bound in Theorem~1 was derived in \cite[Theorem 22.1]{Gamal03}. This completes the converse.

\subsection{Broadcast Channels with Confidential Messages}
Broadcast channels with confidential messages (BCC) \cite{Csiszar02} are wire-tap channels with common messages.
\begin{figure}[t!]
\centering
\psfrag{alice}[][][1]{\text{Joey}}
\psfrag{bob}[][][1]{\text{Chandler}}
\psfrag{eve}[][][1]{\text{Ross}}
\psfrag{M0}[][][1]{$M$}
\psfrag{M1}[][][1]{$M_0$}
\psfrag{W}[][][1]{$W$}
\psfrag{unW}[][][1]{$U^n$}
\psfrag{Xn}[][][1]{$X^n$}
\psfrag{Yn}[][][1]{$Y^n$}
\psfrag{Zn}[][][1]{$Z^n$}
\psfrag{hatM0}[][][1]{$\hat M$}
\psfrag{hatM1}[][][1]{$\hat M_0$}
\psfrag{hatM2}[][][1]{$\wt M_0$}
\psfrag{Pxyz}[][][1]{$Q^n_{YZ|X}$}
\psfrag{f}[][][1]{Encoder}
\psfrag{d1}[][][1]{Decoder $1$}
\psfrag{d2}[][][1]{Decoder $2$}
\psfrag{info}[][][1]{$I(M; Z^n)$}
\includegraphics[width=8.8cm]{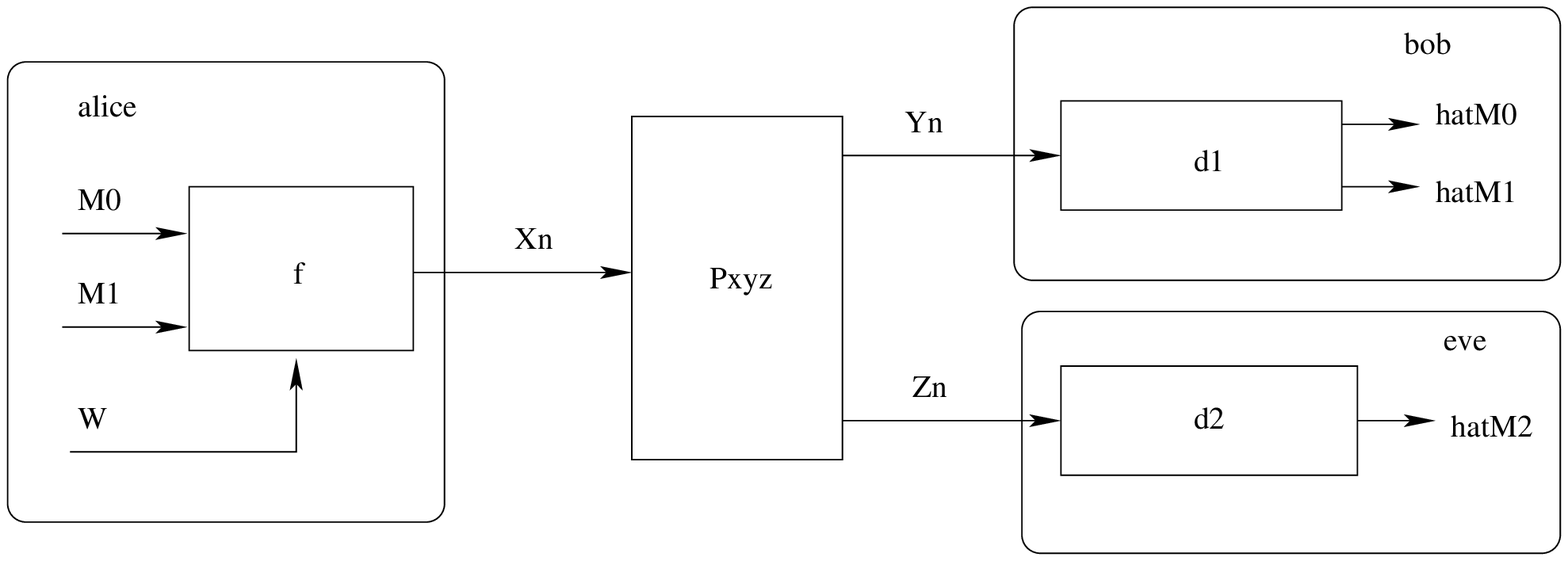}
\caption{A broadcast channel with a confidential message.}
\label{BCC}
\end{figure}
For the BCC (Fig.~\ref{BCC}), Joey has a common message $M_0$ destined for both Chandler and Ross which is independent of
$M$ and uniformly distributed over $\{1,\dots, L_0\}$, $L_0=2^{nR_0}$. An encoder maps $M_0$ and $M$ to
\begin{align}
 X^n=f(M_0, M, W)
\end{align}
which is sent through the channel $Q^n_{YZ|X}$. Chandler estimates $(\hat M_0, \hat M)$ from $Y^n$ while Ross estimates $\wt
M_0$ from $Z^n$. The average error probability is
\begin{align}
P^{\ast(n)}_e=\Pr\left[\left\{(\hat M_0, \hat M)\ne (M_0, M)\right\} \cup \left\{\wt M_0\ne M_0 \right\} \right]
\end{align}
and non-secrecy is measured by $D(P_{MZ^n}||P_MQ^n_Z)$. A rate pair $(R_0, R)$ is achievable if, for any $\xi_1,\xi_2>0$, there is
a sufficiently large $n$, an encoder and two decoders such that
\begin{align}
 P^{\ast (n)}_e& \le \xi_1\\
D(P_{MZ^n}||P_M Q^n_Z)&\le \xi_2.
\end{align}
The effective secrecy capacity region $C_\text{BCC}$ is the closure of the set of achievable $(R_0, R)$. We have the
following theorem.
\begin{theorem}
\label{thm2}
$C_\text{BCC}$ is the same as the weak and strong secrecy capacity region
\begin{align}
C_\text{BCC}=\bigcup
\left\{
\begin{array}{ll}
(R_0, R):\\
0\le R_0 \le \text{min}\;\{ I(U;Y), I(U;Z)\}\\
0\le R \le  I(V;Y|U)-I(V;Z|U)
\end{array}
\right\}
\end{align}
where the union is over all distributions $Q_{UVX}$ satisfying the Markov chain
\begin{align}
&U-V-X-YZ.
\end{align}
One may restrict the alphabet sizes to
\begin{align}
&|\set U|\le |\set X| +3;\quad |\set V|\le |\set X|^2+4 |\set X| +3.
\end{align}
\end{theorem}

\begin{IEEEproof}
The proof is omitted due to the similarity to the proof of Theorem~\ref{thm1}.
\end{IEEEproof}

\subsection{Choice of Security Measures}
Effective secrecy includes both strong secrecy and stealth communication. One may argue that using only $I(M;Z^n)$ or 
$D(P_{Z^n}||Q^n_Z)$ would suffice to measure secrecy. However, we consider two 
examples where secrecy is achieved but not stealth, and where stealth is achieved but not secrecy.
\begin{example}
$I(M;Z^n)\rightarrow 0,\;  D(P_{Z^n}||Q^n_Z)=D> 0$. Suppose that Joey inadvertently uses $\wt Q_X$ rather than $Q_X$ for codebook 
generation, where (\ref{goodcode2}) is still satisfied. The new $\wt Q_X$ could result in a different expected
$\wt Q^n_Z \ne Q^n_Z$. Hence, as $n$ grows large we have
\begin{align}
D(P_{MZ^n}||P_MQ^n_Z)&=I(M;Z^n)+D(\wt Q^n_{Z}||Q^n_Z)
\end{align}
where $I(M;Z^n)\rightarrow 0$ but we have
\begin{align}
D(\wt Q^n_{Z}||Q^n_Z) =D, \;\text{for some}\; D>0.
\end{align}
Ross thus recognizes that Joey is transmitting useful information even though he cannot decode.
\end{example}

\begin{example}
$I(M;Z^n)=I>0,\;  D(P_{Z^n}||Q^n_Z)\rightarrow 0.$

Note that $\text{E}[D(P_{Z^n}||Q^n_Z )]\rightarrow 0$
as $n\rightarrow \infty$ as long as (see \cite[Theorem 1]{Hou01})
\begin{align}
R+R_1>I(X;Z)\label{diver}.
\end{align}
If Joey is not careful and chooses $R_1$ such that (\ref{goodcode2}) is violated and (\ref{diver})
is satisfied, then $D(P_{Z^n}||Q^n_Z)$ can be made small but we have
\begin{align}
I(M;Z^n)=I\;\text{for some}\; I>0.
\end{align}
Thus, although the communication makes $D(P_{Z^n}||Q^n_Z)$ small, Ross can learn
\begin{align}
I(M;Z^n)\approx n[I(X;Z)-R_1]
\end{align}
bits about $M$ if he is willing to pay a price and always tries to decode (see Sec. \ref{Hypo}).
\end{example}

\section{Hypothesis Testing}
\label{Hypo}
The reader may wonder how $D(P_{Z^n}||Q^n_Z)$ relates to stealth. We consider a hypothesis 
testing framework and show that as long as (\ref{diver}) is satisfied, the best Ross can do to detect Joey's action is to guess.
\begin{figure}[t!]
\centering
\psfrag{At}[][][1]{$\set A^n_F$}
\psfrag{cAt}[][][1]{$\left(\set A^n_F\right)^\text{c}$}
\psfrag{T}[][][1]{$\frac{Q^n_Z(z^n)}{P_{Z^n}(z^n)}>F$}
\psfrag{cT}[][][1]{$\frac{Q^n_Z(z^n)}{P_{Z^n}(z^n)}\le F$}
\psfrag{H0}[][][1]{$H_0=Q^n_Z$}
\psfrag{H1}[][][1]{$H_1=P_{Z^n}$}
\includegraphics[width=4.5cm]{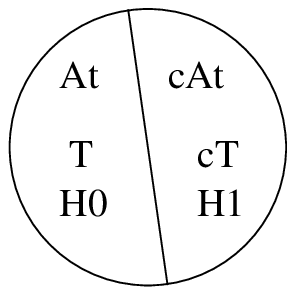}
\caption{Example of the decision regions $\set A^n_F$ and $\left(\set A^n_F\right)^\text{c}$.}
\label{partition}
\end{figure}

For every channel output $z^n$, Ross considers two hypotheses
\begin{align}
&H_0=Q^n_Z\\
&H_1=P_{Z^n}.
\end{align}
If $H_0$ is accepted, then Ross decides that Joey's transmission is not meaningful, whereas if $H_1$
is accepted, then Ross decides that Joey is sending useful messages. We define two kinds of error probabilities
\begin{align}
\alpha&=\Pr\{H_1\;\text{is accepted}\left.\right|H_0\; \text{is true}\}\\
\beta&=\Pr\{H_0\;\text{is accepted}\left.\right|H_1\; \text{is true}\}.
\end{align}
The value $\alpha$ is referred to as {\em the level of significance} \cite{Lehmann01} and corresponds to the probability of
raising a false alarm, while $\beta$ corresponds the probability of mis-detection. In practice, raising a false alarm can be
expensive.
Therefore, Ross would like to minimize $\beta$ for a given tolerance level of $\alpha$. To this end, Ross performs for every 
$z^n$ a ratio test
\begin{align}
\frac{Q^n_Z(z^n)}{P_{Z^n}(z^n)}=r \label{NP}
\end{align}
and makes a decision depending on a threshold $F$, $F\ge 0$, namely
\begin{align}
\left\{
\begin{array}{ll}
H_0\;\text{is accepted}\quad & \text{if}\; r>F\\
H_1\;\text{is accepted}\quad & \text{if}\; r\le F
\end{array}.
\right.
\end{align}
Define the set of $z^n$ for which $H_0$ is accepted as
\begin{align}
\set A^n_F=\left\{z^n:\; \frac{Q^n_Z(z^n)}{P_{Z^n}(z^n)}>F \right\}
\end{align}
and $\left(\set A^n_F\right)^\text{c}$ is the set of $z^n$ for which $H_1$ is accepted (see Fig.~\ref{partition}). Ross
chooses the threshold $F$
and we have
\begin{align}
\alpha&=Q^n_Z(\left(\set A^n_F\right)^\text{c})=1-Q^n_Z\left(\set A^n_F\right)\notag\\
\beta&=P_{Z^n}(\set A^n_F).
\end{align}
The ratio test in (\ref{NP}) is the {\em Neyman-Pearson test} which is {\em optimal} \cite[Theorem 3.2.1]{Lehmann01} in the sense 
that it minimizes $\beta$ for a given $\alpha$. We have the following lemma.

\begin{lemma}
If $D(P_{Z^n}||Q^n_Z)\le \xi_2$, $\xi_2>0$, then with the Neyman-Pearson test we have
\begin{align}
1-g(\xi_2)\le \alpha+\beta \le 1+g(\xi_2) \label{inequ1}
\end{align}
where
\begin{align}
g(\xi_2)=\sqrt{\xi_2\cdot 2\ln2}
\end{align}
which goes to $0$ as $\xi_2 \rightarrow 0$.
\end{lemma}
\begin{IEEEproof}
Since $D(P_{Z^n}||Q^n_Z)\le \xi_2$, we have (see (\ref{VT}))
\begin{align}
||P_{Z^n}-Q^n_Z||_\text{TV}\le \sqrt{\xi_2\cdot 2\ln2}=g(\xi_2)\label{bound1}
\end{align}
where
\begin{align}
||P_X-Q_X||_\text{TV}=\sum_{x\in \set X}\left|P(x)-Q(x)\right| \label{VT}
\end{align}
is the variational distance between $P_X$ and $Q_X$ and where the inequality follows by Pinsker's inequality \cite[Theorem 
11.6.1]{Cover01}. We further have
\begin{align}
&||P_{Z^n}-Q^n_Z||_\text{TV}\notag\\
&=\sum_{z^n\in \set A^n_F}\left|P_{Z^n}(z^n)-Q^n_Z(z^n)\right|\notag\\
&\quad+\sum_{z^n\in \left(\set A^n_F\right)^\text{c}}|P_{Z^n}(z^n)-Q^n_Z(z^n)|\notag\\
&\ge \sum_{z^n\in \set A^n_F}\left|P_{Z^n}(z^n)-Q^n_Z(z^n)\right|\notag\\
&\overset{(a)}{\ge} \left| \sum_{z^n\in \set A^n_F}\left [P_{Z^n}(z^n)- Q^n_Z(z^n)\right]\right|\notag\\
&=\left| P_{Z^n}(\set A^n_F)-Q^n_Z(\set A^n_F)\right|\notag\\
&=\left|\beta-(1-\alpha)\right|\label{bound2}
\end{align}
where (a) follows by the triangle inequality. Combining (\ref{bound1}) and (\ref{bound2}), we have the bounds (\ref{inequ1}).
\end{IEEEproof}
\begin{figure}[t!]
\centering
\psfrag{alpha}[][][0.9]{$\alpha$}
\psfrag{beta}[][][0.9]{$\beta$}
\psfrag{0}[][][0.9]{$0$}
\psfrag{1}[][][0.9]{$1$}
\psfrag{3}[][][0.9]{$1-g(\xi_2)$}
\psfrag{Sa}[][][0.8]{\text{Stealth}}
\psfrag{Vi}[][][0.8]{\text{No Stealth}}
\includegraphics[width=6 cm]{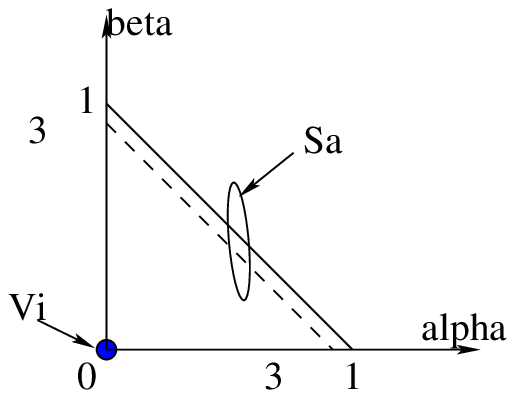}
\caption{Optimal tradeoff between $\alpha$ and $\beta$.}
\label{tradeoff}
\end{figure}
Fig.~\ref{tradeoff} illustrates the optimal tradeoff between $\alpha$ and $\beta$ for stealth communication, i.e., 
when (\ref{diver}) is satisfied. As $n\rightarrow \infty$ and $\xi_2\rightarrow 0$, we have
\begin{align}
D(P_{Z^n}||Q^n_Z)&\rightarrow 0\\
\left(\alpha+\beta\right)&\rightarrow 1.
\end{align}
If Ross allows no false alarm ($\alpha=0$), then he always ends up with mis-detection ($\beta=1$). If
Ross tolerates no mis-detection ($\beta=0$), he pays a high price ($\alpha=1$). Further, for any given
$\alpha$, the optimal mis-detection probability is
\begin{align}
\beta_\text{opt}=1-\alpha.
\end{align}
But Ross does not need to see $Z^n$ or perform an optimal test to achieve $\beta_\text{opt}$. He may
randomly choose some
${\set A}^\prime$ such that
\begin{align}
Q^n_Z((\set A^\prime)^\text{c})=\alpha
\end{align}
and achieves $\beta^\prime_\text{opt}=1-\alpha$. The best strategy is thus to guess. On the other hand, if
\begin{align}
\lim_{n\rightarrow \infty} D(P_{Z^n}||Q^n_Z)>0
\end{align}
then Ross detects Joey's action and we can have
\begin{align}
\alpha+\beta=0.
\end{align}
We thus operate in one of two regimes in Fig. \ref{tradeoff}, either near $(\alpha, \beta)=(0,0)$ or near the line 
$\alpha+\beta=1$.

\section{Discussion}
\label{Diss}

Our resolvability proof differs from that in \cite{Bloch02} in that we rely on {\em unnormalized}
informational divergence \cite{Hou01} instead of variational distance \cite{Verdu01}. Our proof is simpler and the result is 
stronger than that in \cite{Bloch02} when restricting attention to product
distributions and memoryless channels because a small $D(P_{MZ^n}||P_MQ^n_Z)$ implies small $I(M;Z^n)$ and $D(P_{Z^n}||Q^n_Z)$ 
while a small $||P_{X^n}-Q^n_X||_\text{TV}$ implies only a small $I(M;Z^n)$ \cite[Lemma 1]{Csiszar03}.

Hayashi studied strong secrecy for wire-tap channels using resolvability based on unnormalized
divergence and he derived bounds for nonasymptotic cases \cite[Theorem 3]{Ha01}. We remark that Theorem~1 can be derived by
extending \cite[Lemma 2]{Ha01} to asymptotic cases. However, Hayashi did not consider stealth but focused on strong 
secrecy, although he too noticed a formal connection to (\ref{twoTerm}) \cite[p. 1568]{Ha01}.



\section*{Acknowledgment}
J.~Hou and G.~Kramer were supported by an Alexander von Humboldt Professorship endowed by the
German Federal Ministry of Education and Research. G.~Kramer was also supported by NSF Grant CCF-09-05235. J.~Hou thanks
Rafael~Schaefer for useful discussions.

\bibliographystyle{IEEEtran}

\bibliography{bibfile}

\end{document}